\documentclass[a4paper]{article}

\usepackage{INTERSPEECH2022}

\usepackage{booktabs}
\usepackage{multirow}
\usepackage{subcaption}
\usepackage{xcolor}

\usepackage{array}

\title{Probing phoneme, language and speaker information \\in unsupervised speech representations}

\name{Maureen de Seyssel $^{1,2}$, Marvin Lavechin $^{1,3}$, Yossi Adi $^{3}$,\\ Emmanuel Dupoux $^{1,3}$, Guillaume Wisniewski $^2$}

 \address{
   $^1$Cognitive Machine Learning (ENS–CNRS–EHESS–INRIA–PSL Research University), France\\
   $^2$Université de Paris Cité, CNRS, Laboratoire de linguistique formelle, F-75013 Paris, France\\
   $^3$Meta AI Research, France}
 \email{maureen.deseyssel@gmail.com}

\begin{document}

\maketitle

\begin{abstract}
Unsupervised models of representations based on Contrastive Predictive Coding (CPC) \cite{oord2019representation} are primarily used in spoken language modelling in that they encode phonetic information. In this study, we ask what other types of information are present in CPC speech representations. We focus on three categories: phone class, gender and language, and compare monolingual and bilingual models. Using qualitative and quantitative tools, we find that both gender and phone class information are present in both types of models. Language information, however, is very salient in the bilingual model only, suggesting CPC models learn to discriminate languages when trained on multiple languages. Some language information can also be retrieved from monolingual models, but it is more diffused across all features. These patterns hold when analyses are carried on the discrete units from a downstream clustering model. 
However, although there is no effect of the number of target clusters on phone class and language information, more gender information is encoded with more clusters. 
Finally, we find that there is some cost to being exposed to two languages on a downstream phoneme discrimination task.

\end{abstract}

\section{Introduction}

Recent self-supervised models of speech representations capture linguistic features of speech, which can then be used to build language models from raw speech in the context of spoken language modelling \cite{oord2019representation,hsu2021hubert,baevski2020wav2vec}. Specifically, it was found that the output representations of such models encode a significant amount of phonetic information, as suggested by the high scores they yield in phoneme discrimination tasks \cite{dunbar2017zero,nguyen2020zero, millet2022self, chorowski2019unsupervised}. 
However, what is less studied is what other types of information are captured by such acoustic models and how they interact.

In this study, we focus on self-supervised models of speech based on Contrastive Predictive Coding (CPC) \cite{oord2019representation}. Using different probing techniques on their output representations, we want to understand better what information is present and how it is encoded. More specifically, we focus on phonetic, gender and language information. Furthermore, following the growing interest in multilingual representations, we are also interested in how models trained on multiple languages specialise in terms of language information, and for this, we compare models trained on one (monolingual) and two (bilingual) languages.  

Being more aware of the types of information present in such speech representations can be of great interest for downstream applications. Depending on the task, we might want to discard some of this information. In the case of language modelling, we can hope for speaker- and gender-invariant representations, as it is irrelevant to some target tasks. For instance, \cite{van2021analyzing,polyak2021speech,kharitonov2022textless} showed that speaker-specific information is present in CPC-based speech representations, while \cite{van2021analyzing} additionally show that removing it can be helpful in lexical, semantic and syntactic downstream spoken language modelling tasks. 
Yet, other information can prove to be useful. In speech-to-speech translation, for example, multilingual self-supervised models of speech can be used as pretraining to obtain discrete speech units \cite{lee2021textless}. In this context, information about the language can benefit the downstream translation task. Still, while monolingual CPC-based models have shown to transfer well to other languages \cite{kawakami2020learning, riviere2020} in the context of pretraining for Automatic Speech Recognition, no further analyses were done on whether and how language information is present in monolingual and multilingual models.

\begin{figure*}[h]%
\includegraphics[width=\textwidth,keepaspectratio]{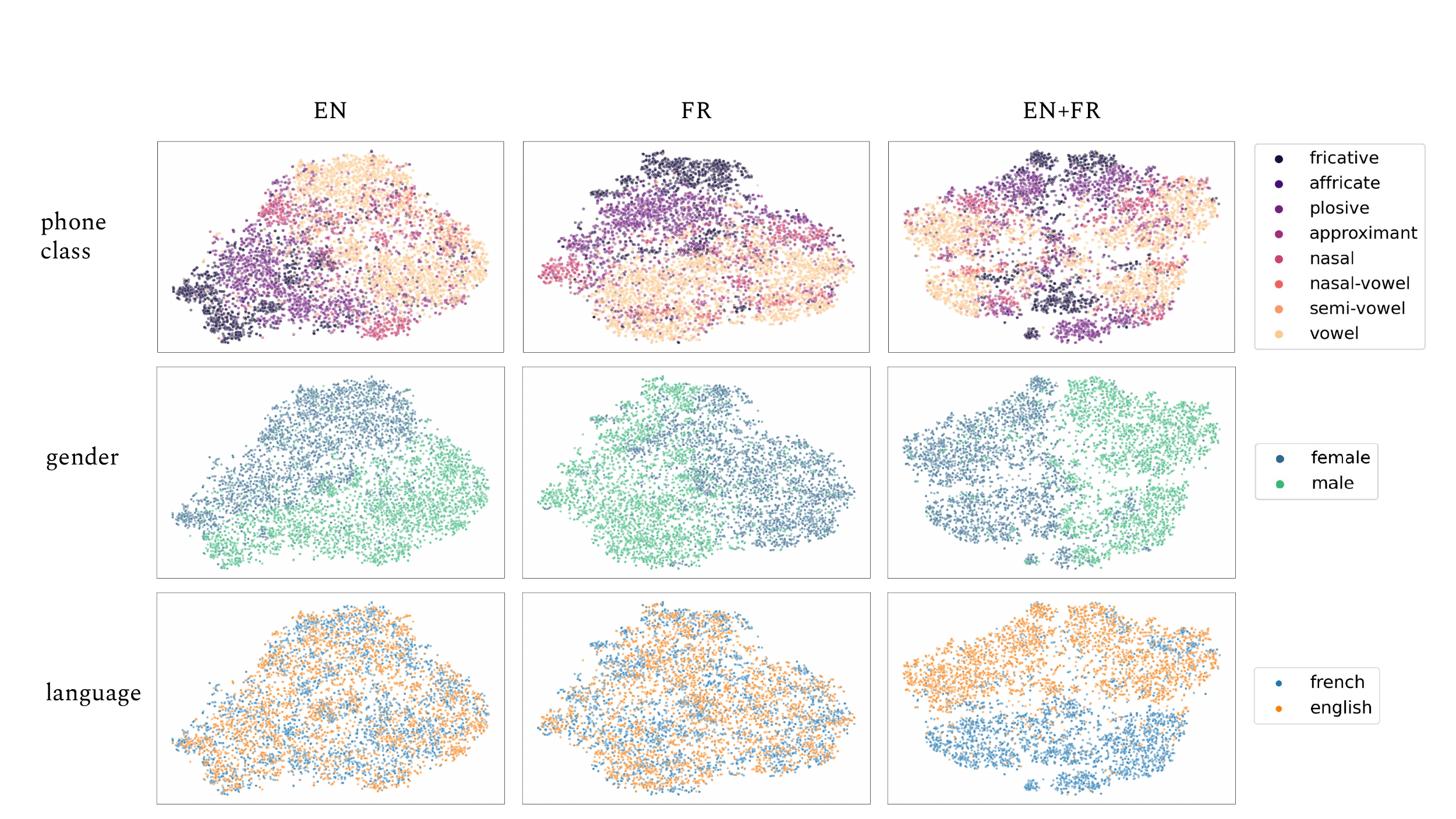}
\caption{T-SNE visualisation of English and French phone embeddings at the CPC level, for monolingual (EN and FR) and bilingual (EN+FR) models. Embeddings are colored based on their phone class label, gender label and language label.}
\label{Figure : tsne-cat}
\end{figure*}

\vspace{0.8em}
\noindent\textbf{Approach.} More precisely, we focus on three categories of information: phonetic class\footnote{We consider the following phonetic classes: fricative; affricate; plosive; approximant; nasal; nasal vowel; semi-vowel; vowel}, gender (male and female), and language (English and French), which are all representations that can be learnt directly from raw data. We also compare two types of models: two monolingual ones (trained exclusively on either English or French) and a bilingual one (trained on both English and French). We first probe the CPC speech representations from these different models for all three categories, using qualitative and quantitative measures (t-SNE visualisation and logistic regression classification). 
Based on past literature, we expect to find phone class and gender-specific information in all models. Whether and how language information is present in the monolingual and bilingual models is more uncertain.
Because in most downstream applications, CPC-based models are followed by a clustering step aiming at transforming the continuous output in discrete units, we will also analyse outputs from clustering k-means algorithms. 
Finally, we will directly look at the differences between monolingual and bilingual models and will test, on a downstream phoneme discrimination task, the impact of having been exposed to two languages.


\section{Experimental setting}

\subsection{Models}

We compare models trained on three conditions: a monolingual English set (EN), a monolingual French set (FR) and a bilingual English and French set (EN+FR). Each train set is made of 3,200h of read speech, retrieved from audiobooks (from the LibriVox and LitteratureAudio projects\footnote{https://librivox.org and http://litteratureaudio.com/} for English and French respectively) and segmented into short utterances using a Voice Activity Detection Model \cite{lavechin2020open}. All three train sets are matched in terms of number of speakers, genre and utterance duration. 
\vspace{-0.1em}

Each train set was used to train an unsupervised acoustic model based on Contrastive Predictive Coding (CPC)\cite{oord2019representation}, using the PyTorch \cite{paszke2019pytorch} implementation from \cite{riviere2020}. During training, this model aims at predicting the near future by selecting the correct frame representation amongst a sample of other negative examples. The architecture and hyperparameters are the same as the CPC-small baseline in \cite{nguyen2020zero}.
Once trained, the model can be used to predict representations of an audio sequence, made of 256 dimensions feature vectors for every 10ms of audio.
\vspace{-0.8em}

\subsection{Evaluation sets}
We downloaded American English and Metropolitan French speech from CommonVoice 7.0 \cite{ardila2019common} to create a 20 hours test set balanced in gender, speakers and languages. We then retrieved for each phone its corresponding audio sequence from the signal, using a phone aligner that we had previously trained using the Kaldi toolkit \cite{povey2011kaldi}. This allowed us to have, for each extracted phone, its audio alignment and its corresponding gender label, language label, and phone class label.

\vspace{-0.5em}

\section{Results}

\subsection{Visualising information in the CPC representations}

For all three models, we extracted the output CPC speech representations of every aligned phoneme from the test set. Because we have a speech representation for every 10ms of speech, we applied a mean pooling function over the speech representations of every frame of a same phoneme to obtain a single 256-dimension feature vector per phoneme. On a subset of the data (N=6,000), we then applied a t-Distributed Stochastic Neighbor Embedding (t-SNE) algorithm to reduce the dimensions from 256 to 2 dimensions \cite{van2008visualizing}, using the Scikit-Learn implementation \cite{sklearn_api}. We plotted the resulting dimensions and colour-coded the data points based on phone class, gender and language for all three models, as presented in Figure \ref{Figure : tsne-cat}.

\begin{table*}[htpb]
\renewcommand{\arraystretch}{1.5}
\caption{Logistic Regression Classification error scores (in \%), on phone class, gender and language, for the EN, FR and EN+FR models. Number of active features is in italics. Inverse $\ell_1$ regularisation strength factor C:  $\ell_1a$ : 0.001; $\ell_1b$ : 0.0001.
}
\label{Table : logReg scores}
\centering
\begin{tabular}{@{}llllllllll@{}}
\toprule
                     & \multicolumn{3}{c}{Phone Class} & \multicolumn{3}{c}{Gender} & \multicolumn{3}{c}{Language} \\
           
          \cmidrule(lr){2-4}
          \cmidrule(lr){5-7}
          \cmidrule(lr){8-10}
                     & EN     & FR     & EN+FR   & EN      & FR     & EN+FR   & EN     & FR     & EN+FR  \\ \midrule

\textbf{LogReg }            &    \textbf{17.6}  \textit{(256)}    &    \textbf{18.0 } \textit{(256)}   &  \textbf{ 15.7 } \textit{(256)}     & \textbf{5.8} \textit{(256)}    & \textbf{6.3 }\textit{(256)}   & \textbf{4.6 }\textit{(256)}    & \textbf{21.7} \textit{(256)}   & \textbf{20.8} \textit{(256)}   & \textbf{8.2} \textit{(256)}   \\ \midrule
LogReg+$\ell_1a$  &  24.7   \textit{(75)}   &    23.7  \textit{(62)}    &    22.1   \textit{(63)}   &  8.2 \textit{(17)}   & 8.8 \textit{(21)}  &  6.4 \textit{(15)}   & 27.7 \textit{(53)}  & 26.6  \textit{(45)}   &   12.0  \textit{(22)}\\
LogReg+$\ell_1b$  &   40.8   \textit{(8)}  &   36.4  \textit{(8)}      &   37.4   \textit{(11)}     & 10.0 \textit{(2)}   & 10.4 \textit{(2)}  & 8.7  \textit{(3)}  & 49.4 \textit{(0)}   & 39.7   \textit{(1)} & 13.2 \textit{(1)}\  \\ 
\bottomrule
\end{tabular}
\end{table*}

First, we note that the phone classes are clearly separated for both the monolingual and bilingual models, suggesting that phonetic information is encoded as expected. Besides, the phone classes are also colour-coded based on how sonorous they are (the darkest ones being the least sonorous), and this sonority seems to be encoded as well, as we can see colour gradients in each model. 
Gender category also seems to be well separated in all three models, as we can see that the two colours are distinct with little overlap. Finally, this qualitative analysis does not show any clear language separation in the monolingual models, whilst there seems to be one in the bilingual model. This can indicate that being exposed to multiple languages leads to encoding some information that can be used to separate languages. More interestingly, visualisation for the bilingual model suggests that language and gender information are clearly distinct and potentially orthogonal to each other.

\subsection{Probing the CPC representations}\label{section : probing_cpc}


While the t-SNE visualisation allows us to get a qualitative idea of how different categories might be separated, it is still necessary to support it with quantitative measures \cite{wattenberg2016how}. Hence, we also use logistic regressions as probes to analyse further what properties are encoded in the speech representations \cite{alain2016understanding}. As in the previous section, we used the phone-level CPC representations, mean-pooled over all frames for each phone. We then split the original test set into an 85/15 train and test set and trained, for each CPC model, three logistic regression models (implementation from Scikit-Learn \cite{sklearn_api}) using either phone classes, gender or language as labels. Error scores are given in the first row of Table \ref{Table : logReg scores}. Note that the scores between phone class and the two other categories are not directly comparable due to the differing number of labels for each category and the unbalanced distribution within the phone class category\footnote{error score achieved by random labelling on the test set: phone class: 76\%; gender: 50\%; language: 50\%}.

Low error scores are reached for both phone class and gender in all three models, supporting the previous results that both phonetic and gender information are encoded in both the monolingual and bilingual models. 
As expected, the language error score for the bilingual model is also very low (8.2\%). Surprisingly, the language error scores for the two monolingual models are also relatively low (21.7\% and 20.8\% for EN and FR, respectively), suggesting that some of the features encoded by the monolingual models can be used to discriminate languages. This result, which was not visible from the t-SNE visualisation, could be due to the fact that the information which is used to reach these low error scores in the monolingual models is diffused within multiple dimensions rather than being strongly encoded in a specific one.

\begin{figure}[htpb]
  \centering
  \includegraphics[width=.9\linewidth]{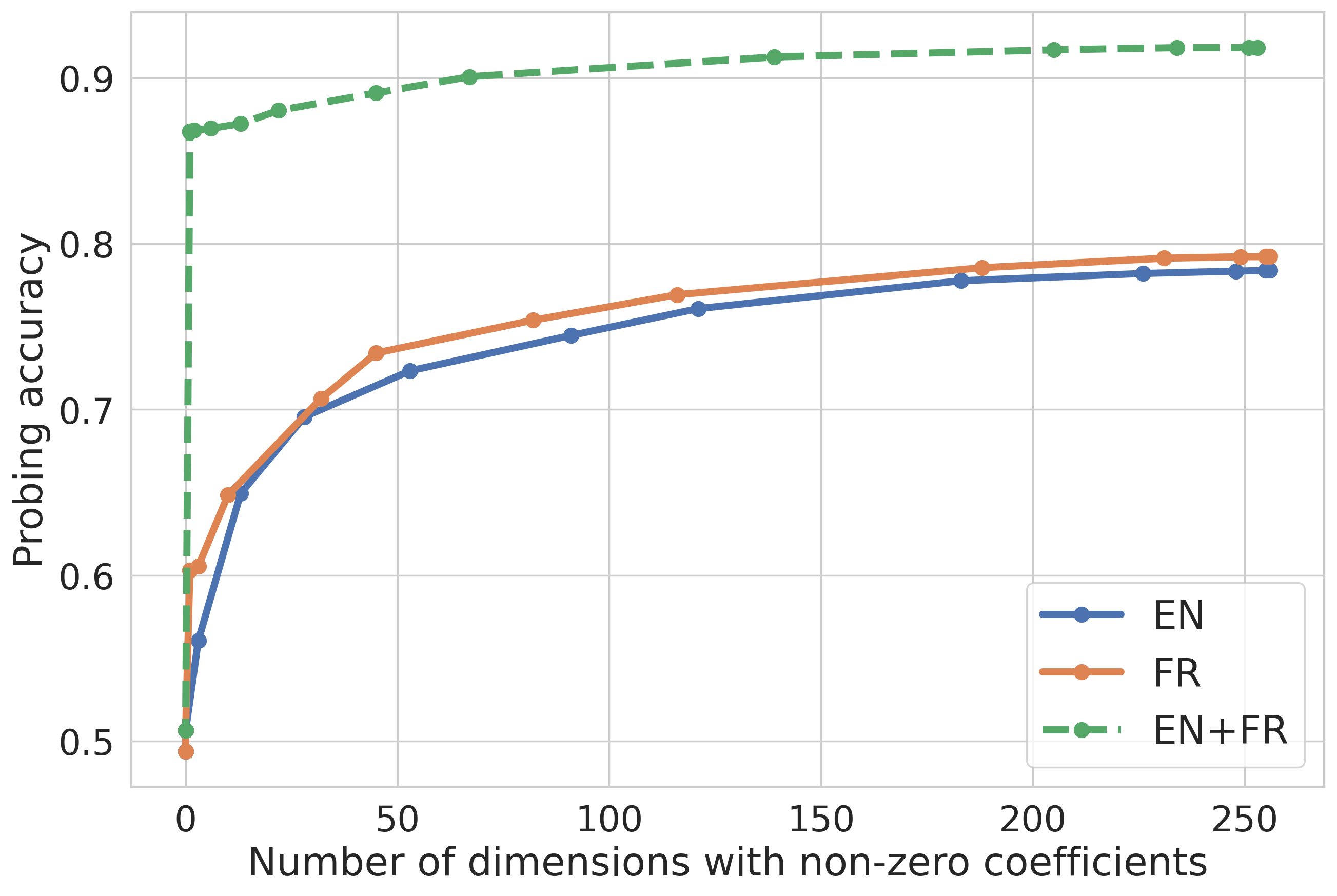}
  \caption{Probing Accuracy on Language Logistic Regression models wrt. number of active coefficients}
  \label{fig:reg_coef}
\vspace{-0.7em}
\end{figure}

To test this hypothesis, we ran other logistic regression analyses using $\ell_1$ regularisation. By doing this, we force the linear model to focus on the most important features, giving less weight to other dimensions. We would expect results with strong $\ell_1$ regularisation to be closer to the pattern seen in the visualisation, with lower language scores in the language category for the monolingual models than for the bilingual one. We tested different strengths of $\ell_1$ regularisation. Error scores along with the number of active features (features that do not have a coefficient of 0 in the logistic regression) are presented in the two last rows of Table \ref{Table : logReg scores}. As expected, the language error scores rise dramatically for the monolingual models compared to the bilingual model when we add $\ell_1$ regularisation. Figure \ref{fig:reg_coef} shows the number of active coefficients in function of accuracy score for the language category. We can see that the monolingual models need to use more features to reach relatively good accuracy scores, compared to the bilingual model, which can already reach 86.8\% accuracy with a single feature. As hypothesised, it indicates that the information used to reach good language accuracy scores in the monolingual models is scattered in a number of dimensions rather than specific to a small number of them.  
For all other categories (and for language in the bilingual model), although the error scores slightly increase, they stay relatively low even with only a very small number of active features, suggesting that there are some specific gender, phone class, and language (for the bilingual model) dimensions. This is also supported by the fact that none of the remaining active features in the model with the strongest $\ell_1$ regularisation overlap between the three categories.

Finally, we can note that the phone class and gender scores for the bilingual model are comparable to those for the monolingual models, suggesting that the additional capture of language information does not prevent the other information to be encoded.

\subsection{Analyses of k-means outputs}

CPC speech representations are used in the context of spoken language modelling. For this purpose, a clustering step is required to transform the continuous features into discrete units before training a language model \cite{nguyen2020zero}. This is why it is also interesting to analyse these discrete units in terms of phone class, gender and language. Therefore, we applied a k-means algorithm with varying number of target clusters (k=50,100,200) on the CPC representations from the EN, FR and EN+FR models. We then retrieved the k-means clusters at each frame of the test set and converted them into one-hot vectors. Finally, for each phone token from the test set, we applied mean-pooling over all its corresponding one-hot vectors to obtain a unique vector of dimension k. Following the method presented in Section \ref{section : probing_cpc}, we trained logistic regression models (without $\ell_1$ regularisation) on each of the phone class, gender and language labels. Classification error scores are presented in Table \ref{Table : logReg scores Kmeans}.   

\begin{table}[htpb]

\caption{Logistic Regression Classification error scores (in \%), on phone class, gender and language, for the EN, FR and EN+FR models. K50 indicates a k-means model of 50 clusters. 
}
\label{Table : logReg scores Kmeans}
\centering
\resizebox{\columnwidth}{!}{%
\begin{tabular}{l rrr rrr rrr}
  \toprule
                     & \multicolumn{3}{c}{Phone Class} & \multicolumn{3}{c}{Gender} & \multicolumn{3}{c}{Language} \\
           
          \cmidrule(lr){2-4}
          \cmidrule(lr){5-7}
          \cmidrule(lr){8-10}
                     & \multicolumn{1}{c}{EN}     & \multicolumn{1}{c}{FR}     & \multicolumn{1}{c}{EN+FR}  & \multicolumn{1}{c}{EN}      & \multicolumn{1}{c}{FR}     & \multicolumn{1}{c}{EN+FR}    & \multicolumn{1}{c}{EN}     & \multicolumn{1}{c}{FR}     & \multicolumn{1}{c}{EN+FR} \\ \midrule

CPC          &    17.6     &    18.0     &   15.7      & 5.8  & 6.3  & 4.6   & 21.7   & 20.8   & 8.2  \\ \midrule
K50 &   29.3      &    28.1   &  30.7 & 22.6      &  24.8  &  22.1  & 40.7    & 41.6      & 25.8   \\
K100 &    28.1    &  27.3     &   28.1    &  16.8  & 20.6   & 15.1   &  39.1 & 40.0  &  25.3 \\
K200 &   27.1     &   25.8     &  25.8     & 14.4  & 18.3  & 12.9   &  37.2  & 38.1   & 24.7 \\

\bottomrule
\end{tabular}
}
\end{table}
\setlength{\textfloatsep}{0.4cm}  

Error scores go up in all three categories and for all models compared to the error scores on the CPC representations, which is expected as the discrete units cannot encode all the information from the continuous representations. Still, phone class and gender classifications reach good scores on the k-means units for both the monolingual and bilingual models. In line with our previous analyses, the bilingual model also reaches much lower language error classification scores on the k-means outputs than the monolingual models, with the latter getting closer to the chance level. This suggests that, for the bilingual model only, the units discovered by the clustering can distinguish the two languages.

\vspace{0.9em}

\noindent\textbf{Effect of number of clusters.} There is little variation on the accuracy scores when changing the number of target clusters for the phone class category with monolingual models, in line with previous studies \cite{nguyen2020zero}. This is also true for the bilingual model, where the effect of number of clusters is very small (16\% error score decrease when considering 200 clusters rather than 50).
There does not seem to be any effect of number of clusters on the language classification either. However, there is an effect of the number of clusters on gender classification, and this for all three models, with an average error score decrease of 35\%  when going from 50 to 200 clusters. This, along with the fact that the logistic regression scores on CPC are lower on gender than on language, suggests that gender is the most present of the three categories within the CPC speech representations.  
Furthermore, as with the results on CPC representations, the bilingual models show comparable discrimination scores to the monolingual ones in both phone class and gender categories, despite having also encoded language information.

\subsection{Comparing monolingual and bilingual models on a phone discrimination downstream task}

Our analyses show that models trained on a single language do not encode directly language-specific information to the same extent that models trained on more languages do. We also found that gender and phone information are still present in bilingual models to the same extent than for the monolingual ones. However, we are also interested in whether there is a cost on downstream tasks to multilingual training. For this, we used the phone ABX task \cite{schatz2013evaluating} to compute discrimination scores for minimal-pairs triplets from the test set. Contrary to the probes used previously, this task is not supervised and allows us to analyse how the representations compare at the phone level without explicitly specifying the relevant features. We tested each model on the language(s) they were trained on (e.g.\ FR models were tested on French triplets, but EN+FR models were tested on English and French triplets). The task was run on both CPC and Kmeans outputs. Within speaker error scores are presented in Table \ref{Table : abxScores}, along with the MFCC baseline scores.
While all three models seem to be able to perform phoneme discrimination from their trained language(s) at a similar level on the CPC representations, the lower ABX scores on k-means clusters for the bilingual model compared to monolingual ones suggest that there is indeed a cost to being exposed to more languages. Furthermore, adding more clusters does not seem to help reduce this difference, discarding the hypothesis that the lower results are due to the larger number of phonemes present in the bilingual model. Understanding where this difference comes from would be of great interest in further studies, especially with the rise of multilingual models in spoken language processing. 

\begin{table}[htpb]
\small
\centering
\caption{Within-speaker on phoneme ABX error scores (in \%). Models are tested on the same language(s) they were trained on. K50 indicates a k-means model of 50 clusters.}
\label{Table : abxScores}
\begin{tabular}{@{}llllll@{}}
\toprule
      & MFCC & CPC & K50  & K100 & K200\\ \midrule
EN    &   17.2   &   9.86  &  17.2   &  18.4 & 19.0  \\
FR    & 17.3      &  11.0   & 19.5   &  19.5  & 19.1\\
EN+FR & 17.3   & 11.7   &  25.3  &  25.1  & 25.6  \\ \bottomrule
\end{tabular}
\end{table}
\vspace{-0.7em}

\section{Discussion \& Conclusions}

The analyses done on CPC speech representations confirm the fact that both phonetic and speaker (using gender as a proxy) information are present in the output features, replicating what was found in the past \cite{nguyen2020zero,van2021analyzing,polyak2021speech,kharitonov2022textless}. Furthermore, these different types of information are still present after converting the CPC continuous representations into discrete units, using a clustering algorithm, even if at a lesser level. This is an important takeaway as different downstream models might need to work on some representations agnostic to one or the other category, depending on their application.

Comparing monolingual and bilingual models, we also found that models trained on multiple languages encode some language information, which seems to be disentangled from gender information. It is not the case for monolingual models, where the language information is less present and more entangled with other features.
Moreover, we found no impact on the quality of phone class and gender features when using bilingual models, with the latter reaching scores comparable to the monolingual models on logistic regression probes. 
Yet, some of our results suggested there is a cost of being trained on multiple languages on a downstream phoneme discrimination task on the discrete units, which cannot be compensated by augmenting the number of target clusters. With multilingual self-supervised models of speech being proposed as pretraining for a series of downstream applications, more analyses of such multilingual representations should be carried out. Another benefit of further studies on the topic would be to understand which cues from the signal (acoustic, phonotactics, prosodic) carry such language information.

Finally, whilst our analyses were carried out on CPC speech representations, the use of the contrastive loss for unsupervised representation learning goes well beyond specifics of our implementation and has been used in numerous works \cite{schneider2019wav2vec, baevski2019vq, baevski2020wav2vec}. Therefore, we think our findings may be applicable to other self-supervised models. In any case, the methodology proposed in this work remains relevant for probing information in any audio representations.



\vspace{0.7em}
\noindent \footnotesize \textbf{Acknowledgments.} MS's work was partly funded by l'Agence de l'Innovation de Défense and performed using HPC resources from GENCI-IDRIS (Grant 20XX-AD011012315). ED in his EHESS role was supported in part by the Agence Nationale pour la Recherche (ANR-17-EURE-0017 Frontcog, ANR-10-IDEX-0001-02 PSL*, ANR-19-P3IA-0001 PRAIRIE 3IA Institute) and a grant from CIFAR (Learning in Machines and Brains).

\bibliographystyle{IEEEtran}

\bibliography{main}

\end{document}